\documentclass[12pt]{article}

\usepackage{cite}
\usepackage{amssymb,amsmath}
\usepackage{epsf}
\usepackage{graphics}

\def\mprp{\mbox{\tiny $\bot$}}
\def\mprl{\mbox{\tiny $\|$}}

\title{\vspace*{-20mm}
\begin{flushright}
{\normalsize Yaroslavl State University\\
             Preprint YARU-HE-02/09\\
             hep-ph/0210029} \\[10mm]
\end{flushright}
General amplitude of the $n$ -- vertex\\
one-loop process in a strong magnetic field}

\author{A. V. Kuznetsov, N. V. Mikheev, D. A. Rumyantsev \\[3mm]
{\small\it Division of Theoretical Physics, 
Yaroslavl State (P.G.~Demidov) University,} \\
{\small\it Sovietskaya 14, 150000 Yaroslavl, Russian Federation}\\
{\small\tt E-mail: avkuzn@uniyar.ac.ru, mikheev@uniyar.ac.ru, rda@uniyar.ac.ru}
}

\date{}

\begin{document}

\maketitle

\thispagestyle{empty}

\begin{abstract}

\baselineskip=18pt

A general analysis of the $n$-vertex loop amplitude in
a strong magnetic field is performed, based on the asymptotic form of
the electron propagator in the field.
As an example, the photon-neutrino processes are considered, where
one vertex in the amplitude is of a general type, and the other vertices are
of the vector type. It is shown, that for odd numbers
of vertices only the $SV_1 \ldots V_{n-1}$ amplitude grows linearly
with the magnetic field strength, while for even numbers of vertices
the linear growth takes place only in the amplitudes $PV_{1} \ldots V_{n-1}$,
$VV_{1} \ldots V_{n-1}$ and $AV_{1} \ldots V_{n-1}$.
The general expressions for the amplitudes of the processes
$\gamma \gamma \to \nu \bar\nu$ (in the framework of the model with the
effective $\nu \nu e e$ -- coupling of a scalar type) and
$\gamma \gamma \to \nu \bar\nu \gamma$ (in the framework of the Standard
Model) for arbitrary energies of particles are obtained.
A comparison with existing results is performed.
\end{abstract}

\vfill

\begin{center}
{\sl Talk presented at the 12th International Seminar \\
``Quarks-2002'', \\
Valday and Novgorod, Russia, June 1-7, 2002}
\end{center}

\newpage

\section{Introduction}

Nowadays, there exists a growing interest to astrophysical objects, where
 the strong magnetic fields with the strength 
$B>B_e$ can be generated $(B_e = m^2/e \simeq 4.41 \cdot 10^{13}$ Gs~\footnote{
We use natural units in which $c = \hbar = 1$, $m$ is the electron mass,  
$e > 0$ is the elementary charge.} is the so-called critical field value). 

The influence of a strong external field on quantum processes is interesting 
because it catalyses the processes, it changes the kinematics and it 
induces new interactions. It is especially important to investigate
the influence of external field on the loop quantum processes where
only electrically neutral particles in the initial and the final states
are presented, such as neutrinos, photons and hypotetical 
axions, familons and so on.
The external field influence on these loop processes is provided by the 
sensitivity of the charged virtual fermion to the field and by the
change of the photon dispersion properties and, therefore,
the photon kinematics.

The research of the loop processes of this type has a rather long history.
The two-vertex loop processes (the photon polarization operator in an 
external field, the decays $\gamma\to\nu\bar{\nu}$, $\nu\to\nu\gamma$ and 
so on) were studed in the
papers~\cite{Tsai:1974,Shabad:88,Skobelev:1995,Gvozdev:1996,Ioannisian:1997}.
The general expression for the two-vertex loop amplitude $j \to f \bar f 
\to j'$ in the homogeneous magnetic and in the crossed field  was obtaned in 
the paper~\cite{Borovkov:1999}, where  all combinations of
scalar, pseudoscalar, vector and axial-vector interactions 
of the generalized currents $j,\,j'$ with fermions were considered. 

A loop process with three vertices is also intresting for theoreticians.
For example, the photon splitting in a magnetic 
field $\gamma \to \gamma \gamma$ is forbidden
in vacuum.    
The review~\cite{Papanian:1986} and the recent
papers~\cite{Adler:1996,Baier:1996,
Baier:1997,Chistyakov:1998,Kuznetsov:1999} were devoted to this process.
One more three-vertex loop process is the conversion of the photon pair 
into the neutrino pair, $\gamma\gamma\to\nu\bar{\nu}$. This process
is interesting as a possible 
channel of stellar cooling. A detailed list of references on 
 this process can be found in our paper~\cite{Kuznetsov:2002a}.

It is well-known (the so-called Gell-Mann theorem~\cite{Gell-Mann:1961}), 
that for massless neutrinos, for both photons on-shell and in the local 
limit of the standard-model weak interaction, the process  
$\gamma\gamma\to\nu\bar{\nu}$ is forbidden.
 Because of this, the four-vertex loop process
with an additional photon $\gamma\gamma\to\nu\bar{\nu}\gamma$ was considered
by some authors. 
In spite of the extra factor $\alpha$, this process has the 
probability larger than the two-photon process.  
The process $\gamma\gamma\to\nu\bar{\nu}\gamma$ was studied both in 
vacuum (from the first paper~\cite{VanHieu:1963} to 
the recent Refs.~\cite{Dicus:1997,
Dicus:1999,Abada:1999a,Abada:1999b,Abada:1999c}), 
and under the stimulating influence of a strong magnetic 
field~\cite{Loskutov:1987,Skobelev:2001,Kuznetsov:2002b}.

So, the calculation of the amplitude of the $n$-vertex loop 
quantum processes ($\gamma \gamma \to \nu \bar\nu$, 
$\gamma \gamma \to \gamma \nu \bar\nu$, 
the axion and familon processes
$\gamma \gamma \to \gamma a$, $\gamma \gamma \to \gamma \varPhi$ 
and so on) in a strong magnetic field is important, because 
these results can be useful for astrophysical applications. 

The paper is organized as follows. A 
general analysis of the $n$-vertex one-loop process amplitude in a strong
magnetic field is performed in Section 2. 
The amplitude, in which the one vertex is of a general type
(scalar $S$, pseudoscalar $P$, vector $V$ or axial-vector $A$), and 
 the other vertices are of the vector type (contracted with photons),
is calculated in Section 3. 
This amplitude is the main result of the paper.
The analitical expressions for the amplitudes of the processes
$\gamma \gamma \to \nu \bar\nu$ and $\gamma \gamma \to \nu \bar\nu \gamma$  
are presented in Sections 4 and 5.

\section{General analysis of the $n$-vertex one-loop processes in a strong 
magnetic field}

We use the effective Lagrangian for the interaction of the generalized 
currents $j$ with electrons in the form:
\begin{eqnarray}
{\cal L}(x) \, = \, \sum \limits_{i} g_i 
[\bar {\psi_e}(x) \Gamma_i \psi_e(x)] j_i,  
\label{eq:L}
\end{eqnarray}
\noindent
where the generic index  $i = S, P, V, A $ numbers the matrices 
$\Gamma_i$, e.g.  
$\Gamma_S = 1, \, \Gamma_P = \gamma_5, \, \Gamma_V = \gamma_{\alpha},
\, \Gamma_A = \gamma_{\alpha} \gamma_5 $,
$j$ is the corresponding quantum object (the current or the photon 
polarisation vector), 
$g_i$ are the coupling constants. 
In particular, for the electron - photon interaction  we have  
$g_V = e,\,\Gamma_V = \gamma_{\alpha},\,j_{V\alpha}(x) = A_\alpha(x)$.

\begin{figure}[htb]

\epsffile[100 541 413 701]{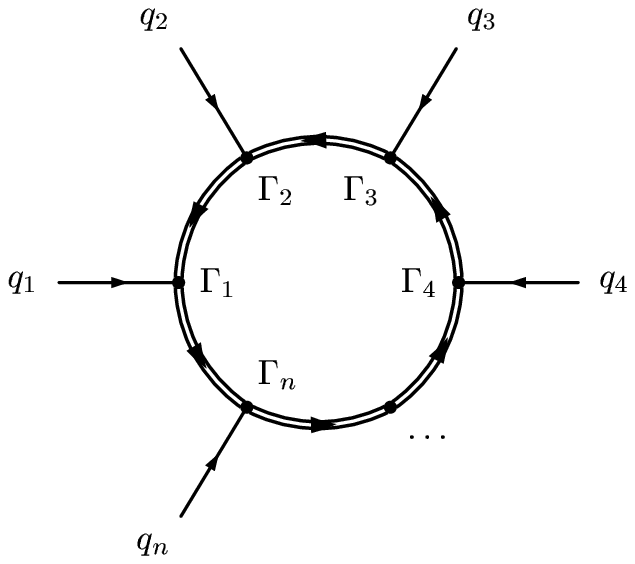}

\vspace{10mm}

\caption{The Feynman diagram for the n-vertex process in a strong magnetic field.}
\label{fig:loop1}

\end{figure}

A general amplitude of the process, corresponding to the effective 
Lagrangian (\ref{eq:L}), is described by fig.~\ref{fig:loop1}.
In the strong field limit, after integration over the coordinates, 
the amplitude takes the form
\begin{eqnarray}
{\cal M}_n \simeq 
 \frac{i \,(-1)^n \,e B}{(2 \pi)^3} 
\exp \! \left (-\frac{R_{\mprp n}}{2eB}\right )
 \int d^2 p_{\mprl} \, \mbox{Tr} \, \big \{\prod \limits^{n}_{k=1} g_k
 \Gamma_k j_k S_{\mprl}(p-Q_k) \big \}, 
\label{eq:M2}
\end{eqnarray}

\noindent 
where $d^2 p_{\mprl} = dp_0 dp_z$, $S_{\mprl}(p) =  \Pi_{-} ((p\gamma)_{\mprl} + m)/(p_{\mprl}^2 - m^2)$
 is the asymptotic form of the electron propagator in the limit  
$eB /\vert m^2 - p_{\mprl}^2 \vert \gg 1$,
$$R_{\mprp 2} = q_{\mprp 1}^2, \quad R_{\mprp 3} = q_{\mprp 1}^2 + 
q_{\mprp 2}^2 + (q_1 \varphi \varphi q_2) + i(q_1 \varphi q_2),$$
$$R_{\mprp n}(n\ge 3) = \sum \limits^{n-1}_{k=1} Q_{\mprp k}^2 -
\sum \limits^{n-1}_{k=2} \sum \limits^{k-1}_{j=1} \left [(Q_k \varphi 
\varphi Q_j) + i(Q_k \varphi Q_j)\right ],$$ 
$$ \quad Q_k = \sum \limits^{k}_{i=1} 
q_i, \quad Q_n = 0,$$
$q_{\mprl}^2 = (q \tilde \varphi \tilde \varphi q)$, 
$q_{\mprp}^2 = (q \varphi \varphi q)$, 
$\varphi_{\alpha \beta} =  F_{\alpha \beta} /B$ 
is the dimensionless field tensor, 
${\tilde \varphi}_{\alpha \beta} = \frac{1}{2} \varepsilon_{\alpha \beta
\mu \nu} \varphi_{\mu \nu}$ is the dual tensor, 
and the indices of the four-vectors and tensors standing inside the 
parentheses are contracted consecutively, e.g.  
 $(a \varphi b) = a_\alpha \varphi_{\alpha \beta} b_\beta$.

As is seen from Eq.~(\ref{eq:M2}), the amplitude depends only on the longitudinal 
components of the momenta, if the magnetic field strength is 
 the maximal physical parameter  
$e B \gg q_{\mprp}^2, q_{\mprl}^2$. 

\section{The photons processes}

Let the vertices $\Gamma_1 \ldots \Gamma_{n-1}$ are of the vector type, and 
the vertex $\Gamma_n$ is arbitrary.
It can be shown that in the limit $q_{\mprp}^2 \ll e B $, for odd numbers
of vertices, only the $S V_1 \ldots V_{n-1}$ amplitude grows linearly
with the magnetic field strength, while for even numbers of vertices
the linear growth takes place only in the amplitudes $PV_{1} \ldots V_{n-1}$,
$VV_{1} \ldots V_{n-1}$ and $AV_{1} \ldots V_{n-1}$.

It should be noted, that in the amplitude~(\ref{eq:M2}) 
the projecting operators $\Pi_-$ separate out  
the photons of only one polarization $(\perp)$ of the two
possible (in Adler's notation~\cite{Ad71})
\begin{eqnarray}
\varepsilon^{(\mprl)}_{\alpha} 
= \frac{\varphi_{\alpha\beta} q_{\beta}}{\sqrt{(q\varphi \varphi q)}},
\qquad 
\varepsilon^{(\mprp)}_{\alpha} 
= \frac{\widetilde \varphi_{\alpha\beta} q_{\beta}}
{\sqrt{(q \widetilde \varphi \widetilde \varphi q)}}.
\label{eq:vect}
\end{eqnarray}

\noindent
As can be deduced from the corresponding analysis, the calculation of 
any type of the amplitude can be reduced to
the evaluation of the scalar integrals
\begin{eqnarray}
S_{n}(Q_{1 \mprl},\ldots ,Q_{n \mprl}) = \int \, \frac{d^2 p_{\mprl}}{(2\pi)^2} \, 
\prod \limits^{n}_{i=1} \frac{1}{(p - Q_i)^2_{\mprl} - m^2 + i\varepsilon}.
\label {eq:s1}
\end{eqnarray}

Notice that the use of the standart method of Feynman parametrization
in calculation of the integrals~(\ref{eq:s1}) can be non-optimal, 
because the number of integrations grows.  
For example, if $n=3$, the double integral~(\ref{eq:s1})
is transformed into the integral over the two Feynman variables. 
If $n=4$, the double integral~(\ref{eq:s1})
is transformed into the integral over the three Feynman variables and so on.
Here we suggest another way. 
By integrating~(\ref{eq:s1}) over $dp_0dp_z$, we obtain
\begin{eqnarray}
S_{n}(Q_{1 \mprl},\ldots ,Q_{n \mprl}) =
 \, \frac{i}{8m^2\pi}
\sum \limits^{n}_{i=1} \underset{l \ne i}{\sum \limits^{n}_{l=1}}
 \, \left [ H \left (\frac{d_{il}^2}{4m^2} \right) + 1 \right ] 
 Re \, \bigg \{ \underset{k \ne i,l}
 {\prod \limits^{n}_{k=1}} \frac{1}{Y_{ilk}} \bigg \} ,   
\label {eq:s4}
\end{eqnarray}
\noindent

where 
\begin{displaymath}
Y_{ilk} = (d_{lk} d_{ik}) \, + \, i (d_{lk} \tilde \varphi d_{ik})
\sqrt{\frac{4m^2}{d_{il}^2} - 1}, \quad 
d_{il}^{\alpha} = Q_{\mprl \, i}^{\alpha} - Q_{\mprl \, l}^{\alpha}.
\end{displaymath}

The function $H(z)$ is defined by the expressions   
\begin{eqnarray}
&&H(z) = \frac{1}{2\sqrt{-z(1 - z)}}
\ln{\frac{\sqrt{1 - z} + \sqrt{-z}}{\sqrt{1 - z} - \sqrt{-z}}} - 1,\quad z<0,
\nonumber \nonumber \\
&&H(z) = \frac{1}{\sqrt{z(1 - z)}}
\arctan{\sqrt{\frac{z}{1 - z}}} - 1,\quad 0<z<1, 
\nonumber \nonumber \\
&&H(z) = - \frac{1}{2\sqrt{z(z - 1)}}
\ln{\frac{\sqrt{z} + \sqrt{z - 1}}{\sqrt{z} - \sqrt{z - 1}}}
 - 1 + \frac{i\pi}{2\sqrt{z(z - 1)}}  ,\quad z>1, 
\nonumber 
\end{eqnarray}
\noindent

and it has the asymptotics
\begin{eqnarray}
&&H(z) \simeq \frac{2}{3}z + \frac{8}{15}z^2 + \frac{16}{35}z^3, \quad |z| \ll 1,
\label {eq:s5}
\end{eqnarray}
\begin{eqnarray}
&&H(z) \simeq -1 - \frac{1}{2z}\ln{4|z|}, \quad |z| \gg 1.
\label {eq:s6}
\end{eqnarray}

\section{The process $\gamma \gamma \to \nu \bar\nu$}

Let us apply the results obtained to the calculation of some 
quantum processes. For the amplitude of the process 
$\gamma \gamma \to \nu \bar\nu$ in the framework of the model with the
effective $\nu \nu e e$ -- coupling of a scalar type we obtain from 
Eqs.~(\ref{eq:M2}),~(\ref{eq:s1}),~(\ref{eq:s4}) 
\begin{eqnarray}
{\cal M}_{3}^s \,& = &\, 
 \frac{2\alpha}{\pi} \, \frac{B}{B_e}  \, g_{s} \, j_s \, m   
 \, \frac{(q_1 \widetilde \varphi \varepsilon^{(1)})
 (q_2 \widetilde \varphi \varepsilon^{(2)})}
 {4m^2 [(q_1 q_3)^2_{\mprl} - q^2_{1\mprl} q^2_{3\mprl}] +
 q^2_{1\mprl} q^2_{2\mprl} q^2_{3\mprl}} \times
\nonumber \nonumber \\
&\times &\, \left \{ \left [q^2_{1\mprl} q^2_{3\mprl} -
2m^2 (q^2_{3\mprl} + q^2_{1\mprl} - q^2_{2\mprl})\right ] 
H \left (\frac{q_{1\mprl}^2}{4m^2} \right ) \right. \, + 
\nonumber \nonumber \\
& + & \, \left [q^2_{2\mprl} q^2_{3\mprl} -
2m^2 (q^2_{3\mprl} + q^2_{2\mprl} - q^2_{1\mprl})\right ]
H \left (\frac{q_{2\mprl}^2}{4m^2} \right ) \, + 
\nonumber \nonumber \\
& + & \, \left. q^2_{3\mprl}(4m^2 -q^2_{3\mprl})
H \left (\frac{q_{3\mprl}^2}{4m^2} \right ) -
2q^2_{3\mprl}(q_1q_2)_{\mprl} \, \right \}, 
\label{eq:MS3}
\end{eqnarray}

\noindent
where $g_s = - 4 \; \zeta \; G_{\mbox{\normalsize{F}}}/\sqrt{2}$
is the effective $\nu \nu e e$ -- coupling constant in the 
left-right-symmetric extension of the Standard Model,  
$\zeta$ is the small mixing angle of the charged $W$  bosons, 
$j_s = [\bar \nu_e(p_1) \nu_e(-p_2)]$ is Fourier transform
of the scalar neutrino current, $q_3 = p_1 + p_2$
is the neutrino pair momentum.

Substituting the photon polarization vector $\varepsilon^{(\mprp)}_{\alpha}$ 
 from Eq.~(\ref{eq:vect}) into~(\ref{eq:MS3})
and using~(\ref{eq:s5}) and~(\ref{eq:s6}), we obtain the asymptotics:

\begin{itemize}
\item[a)] 
at low photon energies, $\omega_{1,2} \lesssim m$
\begin{eqnarray} 
{\cal M}_{3}^{s} \simeq \frac{8 \alpha}{3 \pi} \,
\frac{G_{\mbox{\normalsize{F}}}}{\sqrt{2}}\,
\frac{\zeta}{m}\;\frac{B}{B_e}\,
\left [\bar\nu_e (p_1) \, \nu_e (- p_2) \right ]\,
\sqrt{q_{1\mprl}^2 q_{2\mprl}^2} ;
\label{eq:M<}
\end{eqnarray}

\item[b)] 
at high photon energies, $\omega_{1,2} \gg m$, in the leading log 
approximation:
\begin{eqnarray} 
{\cal M}_{3}^{s} \simeq \frac{16 \alpha}{\pi} \,
\frac{G_{\mbox{\normalsize{F}}}}{\sqrt{2}}\,
\zeta\;\frac{B}{B_e}\,m^3 \,
\left [\bar\nu_e (p_1) \, \nu_e (- p_2) \right ]\,
\frac{1}{\sqrt{q_{1\mprl}^2 q_{2\mprl}^2}}\,
\ln \frac{\sqrt{q_{1\mprl}^2 q_{2\mprl}^2}}{m^2}.
\label{eq:M>}
\end{eqnarray}

\end{itemize}
These expressions coincide with the results obtained in the 
paper~\cite{Kuznetsov:2002a}.

\section{The process $\gamma \gamma \to \nu \bar\nu \gamma $}

The process of this type, where one initial photon is virtual, namely, the  
 photon conversion into neutrino pair on nucleus
 was considered, in the framework of the Standard Model, 
in the papers~\cite{Skobelev:2001, Kuznetsov:2002b}. 
This process can be studied by using the amplitude of the transition
$\gamma \gamma \to \nu \bar\nu \gamma $, which can be obtained from 
Eq.~(\ref{eq:M2}) in the form: 
\begin{eqnarray}
{\cal M}_{4}^{VA} \,& = &\, - \, 
\frac{8 i e^3}{\pi^2} \, \frac{B}{B_e}  \,
\frac{G_{\mbox{\normalsize{F}}}}{\sqrt{2}} \, m^2 \times   
\nonumber \nonumber \\
&\times & (q_1 \widetilde \varphi \varepsilon^{(1)})
 (q_2 \widetilde \varphi \varepsilon^{(2)})
 (q_3 \widetilde \varphi \varepsilon^{(3)})
[g_V (j\widetilde \varphi q_4) +
 g_A (j\widetilde \varphi \widetilde \varphi q_4)] \times
\nonumber \nonumber \\
&\times & \frac{1}{D} \left \{ I_4(q_{1\mprl},q_{2\mprl},q_{3\mprl}) 
+ I_4(q_{2\mprl},q_{1\mprl},q_{3\mprl}) + 
I_4(q_{1\mprl},q_{3\mprl},q_{2\mprl}) \right \},
\label{eq:MVA4}
\end{eqnarray}

\noindent
where
$g_V,\;g_A$ are the vector and axial-vector constants of the effective 
$\nu \nu e e$  Lagrangian of the Standard Model, 
$g_V = \pm 1/2 + 2 \sin^2 \theta_W, \; g_A = \pm 1/2$, 
here the upper signs correspond to the electron neutrino, and   
lower signs correspond to the muon and tau neutrinos;
$j_{\alpha} = [\bar \nu_e(p_1) \gamma_{\alpha} (1 + \gamma_5) \nu_e(-p_2)]$
 is the Fourier transform of the neutrino current;
 $q_4 = p_1 + p_2$  is the neutrino pair momentum;
$$D = (q_1q_2)_{\mprl}(q_3q_4)_{\mprl} + (q_1q_3)_{\mprl}(q_2q_4)_{\mprl} +
(q_1q_4)_{\mprl}(q_2q_3)_{\mprl}.$$

The formfactor $I_4 (q_{1 \mprl} ,q_{2 \mprl} ,q_{3 \mprl})$ is expressed 
in terms of the integrals~(\ref{eq:s1}),~(\ref{eq:s4})  
\begin{eqnarray}
&&I_4 (q_{1 \mprl}, q_{2 \mprl}, q_{3 \mprl}) \, =  \, 
S_3 (q_{1 \mprl} + q_{2 \mprl}, q_{4 \mprl},0) + S_3 (q_{1 \mprl}, q_{4 \mprl},0) +
\nonumber \nonumber \\
&& + S_3 (q_{1 \mprl} + q_{2 \mprl}, q_{1 \mprl},0) +
 S_3 (q_{2 \mprl} - q_{3 \mprl}, q_{2 \mprl},0) +
\nonumber \nonumber \\
&& + [6 m^2 - (q_1 + q_2)_{\mprl}^2 - (q_2 - q_3)_{\mprl}^2]
S_4 (q_{1 \mprl}, q_{1 \mprl}+q_{2 \mprl}, q_{4 \mprl},0). 
\end{eqnarray}

Using the asymptotics of the functions $H(z)$, we obtain
\begin{itemize}
\item[a)] 

at low photon energies, $\omega_{1,2,3} \ll m$
\begin{eqnarray}
&&{\cal M}_{4}^{VA} \, \simeq \,  
 - \frac{2 e^3}{15\pi^2} \, \frac{B}{B_e}  \, 
\frac{G_{\mbox{\normalsize{F}}}}{\sqrt{2}} \, \frac{1}{m^4} \times  
\nonumber \nonumber \\
&&\times (q_1 \widetilde \varphi \varepsilon^{(1)})
 (q_2 \widetilde \varphi \varepsilon^{(2)})
 (q_3 \widetilde \varphi \varepsilon^{(3)})
 [g_V (j\widetilde \varphi q_4) +
 g_A (j\widetilde \varphi \widetilde \varphi q_4)],
\label{eq:I_4_low}
\end{eqnarray}

which is in agreement with the result of the 
paper~\cite{Kuznetsov:2002b};
\item[b)] 
at high photon energies, $\omega_{1,2,3} \gg m$, in the leading log 
approximation we obtain:
\begin{eqnarray}
&&{\cal M}_{4}^{V,A} \, \simeq \,
 - \frac{8 e^3}{3\pi^2} \, \frac{G_{\mbox{\normalsize{F}}}}{\sqrt{2}} \,
\frac{B}{B_e}  \, m^4 \times
\nonumber \nonumber \\
&&\times (q_1 \widetilde \varphi \varepsilon^{(1)})
 (q_2 \widetilde \varphi \varepsilon^{(2)})
 (q_3 \widetilde \varphi \varepsilon^{(3)})
 [g_V (j\widetilde \varphi q_4) +
 g_A (j\widetilde \varphi \widetilde \varphi q_4)] \times
\nonumber \nonumber \\
&&\times \frac{1}{q^2_{1 \mprl}q^2_{2 \mprl}q^2_{3 \mprl}q^2_{4 \mprl}} 
 \ln \frac{\sqrt{q_{1 \mprl}^2 q_{2 \mprl}^2 q_{3 \mprl}^2}}{m^3}.
\label{eq:I_4_hig}
\end{eqnarray}
\end{itemize}
To the best of our knowledge, this result is obtained for the first time.

\section{Conclusions}

We have obtained the general expressions~(\ref{eq:MS3}) and~(\ref{eq:MVA4}) for
the amplitudes of the processes
$\gamma \gamma \to \nu \bar\nu$ (in the framework of the model with the
effective $\nu \nu e e$  coupling of a scalar type) and 
$\gamma \gamma \to \nu \bar\nu \gamma$ (in the framework of the Standard Model)
for arbitrary energies of particles. 
A comparison with the existing results has been performed.

\bigskip

{\bf Acknowledgements}  

We express our deep gratitude to the organizers of the 
Seminar ``Quarks-2002'' for warm hospitality.

This work was supported in part by the Russian Foundation for Basic 
Research under the Grant No.~01-02-17334
and by the Ministry of Education of Russian Federation under the 
Grant No.~E00-11.0-5.

%%%%%%%%%%%%%%

%%%%%%%%%%%%%%%%%%%%%%%%%%%%%%%%%%%%%%%%%%%%%%%%%%%%%%%%%%%%%%%%%%%%%%%%%
\end{document}